\documentclass[10pt,twocolumn,letterpaper]{article}

\usepackage{iccv}              

\definecolor{iccvblue}{rgb}{0.21,0.49,0.74}
\usepackage[pagebackref,breaklinks,colorlinks,allcolors=iccvblue]{hyperref}
\usepackage{makecell, tabularx, booktabs, xcolor}
\usepackage{float}
\usepackage{multirow}
\usepackage{colortbl}  
\usepackage[accsupp]{axessibility}  

\title{CABLD: Contrast-Agnostic Brain Landmark Detection with Consistency-Based Regularization}

\author{
    Soorena Salari$^{1}$\thanks{Correspondence to: \texttt{soorena.salari@concordia.ca}.} \hfill
    Arash Harirpoush$^{1}$ \hfill
    Hassan Rivaz$^{2}$ \hfill
    Yiming Xiao$^{1}$ \\
    $^1$Department of Computer Science and Software Engineering, Concordia University, Montréal, Canada \\
    $^2$Department of Electrical and Computer Engineering, Concordia University, Montréal, Canada
}

\begin{document}
\maketitle
\begin{abstract}
Anatomical landmark detection in medical images is essential for various clinical and research applications, including disease diagnosis and surgical planning. However, manual landmark annotation is time-consuming and requires significant expertise. Existing deep learning (DL) methods often require large amounts of well-annotated data, which are costly to acquire. In this paper, we introduce \textbf{CABLD}, a novel self-supervised DL framework for 3D brain landmark detection in unlabeled scans with varying contrasts by using only a single reference example. To achieve this, we employed an inter-subject landmark consistency loss with an image registration loss while introducing a 3D convolution-based contrast augmentation strategy to promote model generalization to new contrasts. Additionally, we utilize an adaptive mixed loss function to schedule the contributions of different sub-tasks for optimal outcomes. We demonstrate the proposed method with the intricate task of MRI-based 3D brain landmark detection. With comprehensive experiments on four diverse clinical and public datasets, including both T1w and T2w MRI scans at different MRI field strengths, we demonstrate that \textbf{CABLD} outperforms the state-of-the-art methods in terms of mean radial errors (MREs) and success detection rates (SDRs). Our framework provides a robust and accurate solution for anatomical landmark detection, reducing the need for extensively annotated datasets and generalizing well across different imaging contrasts. Our code is publicly available at \href{https://github.com/HealthX-Lab/CABLD}{https://github.com/HealthX-Lab/CABLD}.
\end{abstract}    
\section{Introduction}
\label{sec:intro}

Anatomical landmarks in medical images are salient keypoints with clinical significance in bodily structures, such as the knee \cite{kordon2019multi}, spine \cite{yi2020vertebra}, and lungs \cite{urschler2006automatic}. They often serve as essential reference points to facilitate anatomical navigation and measurement of physiological changes for diagnostic purposes and surgical planning. With an aging global population, neurological disorders are becoming increasingly prominent issues. With a long history in neurological research and patient care, brain landmarks as shown in MRI are of great interest to the clinical and research communities, but are more complex due to the intricate neuroanatomy, softer geometric features (compared to bones), and high-precision requirement in deployment. Besides the traditional task of anatomical localization for planning functional neurosurgery \cite{xiao2020image}, brain landmarks \cite{zhang2017detecting,taha2023magnetic} have also been employed to diagnose neurodegenerative conditions, such as Parkinson’s disease \cite{abbass2022application} and Alzheimer’s disease \cite{zhang2017alzheimer,liu2018landmark,liu2020anatomical,liu2018anatomical,zhang2017landmark}. Additionally, for big data analysis of brain MRIs, where subject-to-template image registration is crucial, brain landmarks can provide the essential tool to assess the quality of image alignment to ensure the reliability of the study. Notably, thanks to their high spatial specificity, brain landmarks can offer better sensitivity and interpretability in registration quality assessment than other volume-overlap-based metrics (e.g., Dice coefficient) \cite{rohlfing2011image}. However, manual landmark annotation requires significant domain expertise, time, and labor. Thus, automated anatomical landmark detection is highly instrumental.

To date, a number of deep learning (DL)-based techniques \cite{chen2021fast, zhang2017detecting, mccouat2022contour} have been proposed for accurate and efficient anatomical landmark detection in the forms of heatmaps \cite{tompson2014joint} or regressed coordinates \cite{salari2023uncertainty}. However, three main challenges remain in this domain. \textbf{First}, due to aforementioned challenges, very few public datasets and protocols for anatomical landmark detection exist, particularly for 3D brain MRI, posing major barriers to the development and validation of relevant DL-based methods. \textbf{Second}, to mitigate the impact of data scarcity, label-efficient methods \cite{zhu2023uod,quan2022images,quan2024images} have gained great interest, particularly with recent adoption of self-supervised learning (SSL) to generate anatomical landmarks by training keypoint-based image registration models \cite{bhalodia2020self,bhalodia2021leveraging,evan2022keymorph,wang2023robust,wang2024brainmorph}. However, these methods often fail to produce landmarks with subject-consistent localization protocols, which must consider both geometric saliency and anatomical/physiological significance, thus greatly limiting their clinical applicability. \textbf{Finally}, sensitivity to scanner types, scanning parameters, and image contrasts has been a well-known issue in DL-based medical image analysis, including automatic anatomical landmark detection. Besides regular Gamma-correction-based contrast augmentation, existing landmark detection techniques have attempted to mitigate this issue by adopting computationally expensive metrics, such as Mutual Information (MI) and Modality Independent Neighborhood Descriptor (MIND) \cite{bhalodia2021leveraging} and incorporating additional loss functions to enforce cross-contrast consistency \cite{wang2023robust}. However, these measures still require real multi-contrast datasets during training, which could be difficult to obtain. To mitigate this, contrast synthesis has also been proposed by using anatomical segmentation maps \cite{hoffmann2021synthmorph}, but these dense segmentation maps may not be readily available and may struggle to capture nuanced details in scans. Alternatively, more flexible and drastic contrast alteration based on CNN \cite{sideri2023mad,wang2024car,ouyang2022causality} may offer the needed benefits, but has not been explored for volumetric data. 

As implied by recent works \cite{bhalodia2020self,bhalodia2021leveraging,evan2022keymorph,wang2023robust,wang2024brainmorph,chao2023self}, the tasks of image registration and anatomical landmark localization learning share a strong synergy. Such duality may offer a more elegant and efficient solution for accurate landmark tagging than conventional hierarchical strategies \cite{chen2021fast, lang2019automatic}. This motivates us to adopt a multi-task learning approach, which carefully considers the balance of the learning schedules for these two tasks \cite{lian2020multi, browatzki20203fabrec} for our goal. To address the limitations of previous works, we present a novel self-supervised DL framework, called \textbf{CABLD}, that employs multi-task learning for contrast-agnostic anatomical landmark detection. With unbiased representation of anatomical features, population-averaged anatomical atlases made from medical scans of a large cohort \cite{xiao2023population} are often used as the reference for spatial normalization and anatomical navigation in clinical tasks. We hypothesize that they can provide the means to inject needed expert knowledge into an annotation-efficient SSL framework for landmark detection. We demonstrate our method for the challenging task of 3D brain landmark detection in multi-contrast MRIs, which have been largely under-explored despite the great demand. The major contributions of this paper are as follows:

    \begin{itemize} 

    \item We introduce an innovative and data-efficient self-supervised framework for brain landmark detection by leveraging inter-scan landmark consistency to enforce a systematic landmark protocol across subjects, with the requirement of only a single reference atlas annotation. 
    
    \item To enable model generalization to unseen MRI contrasts, we propose an effective 3D contrast augmentation strategy using random convolution without the need for real multi-contrast training data. 

    \item We propose an adaptive mixed loss function, inspired by curriculum learning, to dynamically balance registration similarity and landmark consistency for robust anatomical feature learning.

    \item We conduct comprehensive validation on the proposed method using four diverse clinical and public datasets at different MR field strengths. The results demonstrate the efficacy of the proposed components, achieving significant performance improvements over SOTA approaches. 

    \item We demonstrate our method's clinical applicability in exploratory tasks of neurological disease diagnosis.

    \end{itemize}

\section{Related Work}
\label{sec:relatedwork}

\subsection{Landmark Transfer via Atlas Propagation}

Classical atlas-to-subject registration methods (e.g., ANTs \cite{avants2009advanced} and NiftyReg \cite{modat2010fast}) are foundational approaches for anatomical landmark detection, relying on the calculation of deformation vector fields (DVFs) to warp landmarks from the atlas space to the subject space. While effective for global alignment, these methods are slow, computationally expensive, and often fail to capture fine-grained anatomical details, limiting their utility in time-sensitive and precision-critical applications. Although more recent DL-based registration can significantly enhance inference speed, classic methods still set the bar for accuracy \cite{chen2024survey}. However, atlas-propagation-based landmarks are sub-optimal for registration quality assessment due to the inherent bias toward the source registration method.

\subsection{Supervised Landmark Detection} 
Supervised DL methods for landmark detection have dominated the literature by generating heat maps \cite{mccouat2022contour} or regressing the position coordinates of the predicted keypoints \cite{zhang2017detecting,li2020end}. To reduce the computational cost of these models for 3D images, some works also reformat the task into 2.5D feature extraction instead \cite{yun2020learning}. Multi-stage techniques, such as 3D Faster R-CNN \cite{chen2021fast} and Mask R-CNN \cite{lang2019automatic}, identify landmarks within initially localized target regions in a coarse-to-fine paradigm. Also, U-Net architectures, combined with attention mechanisms, have been frequently employed to improve the accuracy of anatomical landmark detection \cite{lang2020automatic,li2020end}. Nevertheless, the primary bottleneck of supervised landmark detection is the poor availability of annotated landmarks, thus motivating novel SSL frameworks to tackle such challenges.

\subsection{Self-Supervised Landmark Detection}
Recently, SSL has gained interest in anatomical landmark detection. Chao et al. \cite{chao2023self} propose a framework that discovers anatomically consistent landmarks between scans by combining a landmark discovery loss with a deformation reconstruction loss, leveraging an external registration model to ensure accurate correspondences. However, the accuracy of their method can heavily depend on the chosen registration model, with potential model biases and limited adaptability to unseen contrasts. Medical image registration frameworks that leverage keypoint alignment between scans have also been explored for landmark discovery to help visually explain DL-based registration outcomes \cite{wang2023robust,wang2024brainmorph,bhalodia2021leveraging,bhalodia2020self,wang2023robust}. These techniques extract landmarks from two subjects and subsequently estimate the DVFs. However, as the main objective of these methods is image registration, with the primary guidance of conventional similarity losses, such as MSE and MIND (along with DVFs smoothness constraints), the ``discovered" landmark pairs are often located in regions that are strategic for image alignment and vary among inference samples. Thus, it is not ideal for generating clinical anatomical landmarks, which follow consistent protocols across subjects and image contrasts. Finally, to enable adaptability to different image contrasts, Wang et al. \cite{wang2023robust} presented a self-supervised loss to enforce consistency between predicted landmarks for the same subject's multi-contrast scans. However, the limited multi-contrast/modal datasets can pose difficulties.

\subsection{Multi-task Learning} 
Multi-task learning is often employed for landmark detection tasks by associating landmark detection with other downstream tasks, such as image classification, image segmentation, and reconstruction \cite{honari2018improving, torosdagli2018deep, zhang2020context}. Browatzk et al. \cite{browatzki20203fabrec} employed supervised transfer learning based on facial image reconstruction for landmark detection. Lian et al. \cite{lian2020multi} introduced a dynamic multi-task Transformer network to detect dental landmarks together with bone segmentation. In our study, we couple the tasks of landmark detection with image registration. Unlike other multi-task learning methods that share feature embeddings, our framework uses image registration to assist the model in initializing the landmarks, after which our consistency regularization terms calibrate the detected landmarks. A curriculum learning approach \cite{wang2021survey,bengio2009curriculum} is also employed to mitigate the interferences between the two tasks.

\section{Materials and Methods}
\label{sec:method}

\subsection{Problem Setup and Overview}
Let \( \boldsymbol{{X}} = \{x_i \subseteq \mathbb{R}^N\}_{i=1,\dots, K} \) be the training set of \( K \) unlabeled volumetric scans. Additionally, we used a single template image, \( I_{\text{template}} \subseteq \mathbb{R}^N \) that contains \( L \) expert-annotated landmarks, \( \textbf{\textit{P}} = \{P_n \in \mathbb{R}^3 \}_{n=1,\dots, L} \). Given the training scans and a single template with annotated landmarks, our objective is to optimize the anatomical landmark detection model \( \boldsymbol{f}(\cdot; \theta) \) to accurately detect the corresponding anatomical landmarks that are defined in the single template in the other scans. 

\begin{figure*}[tb]
\centering
\includegraphics[scale=0.273]{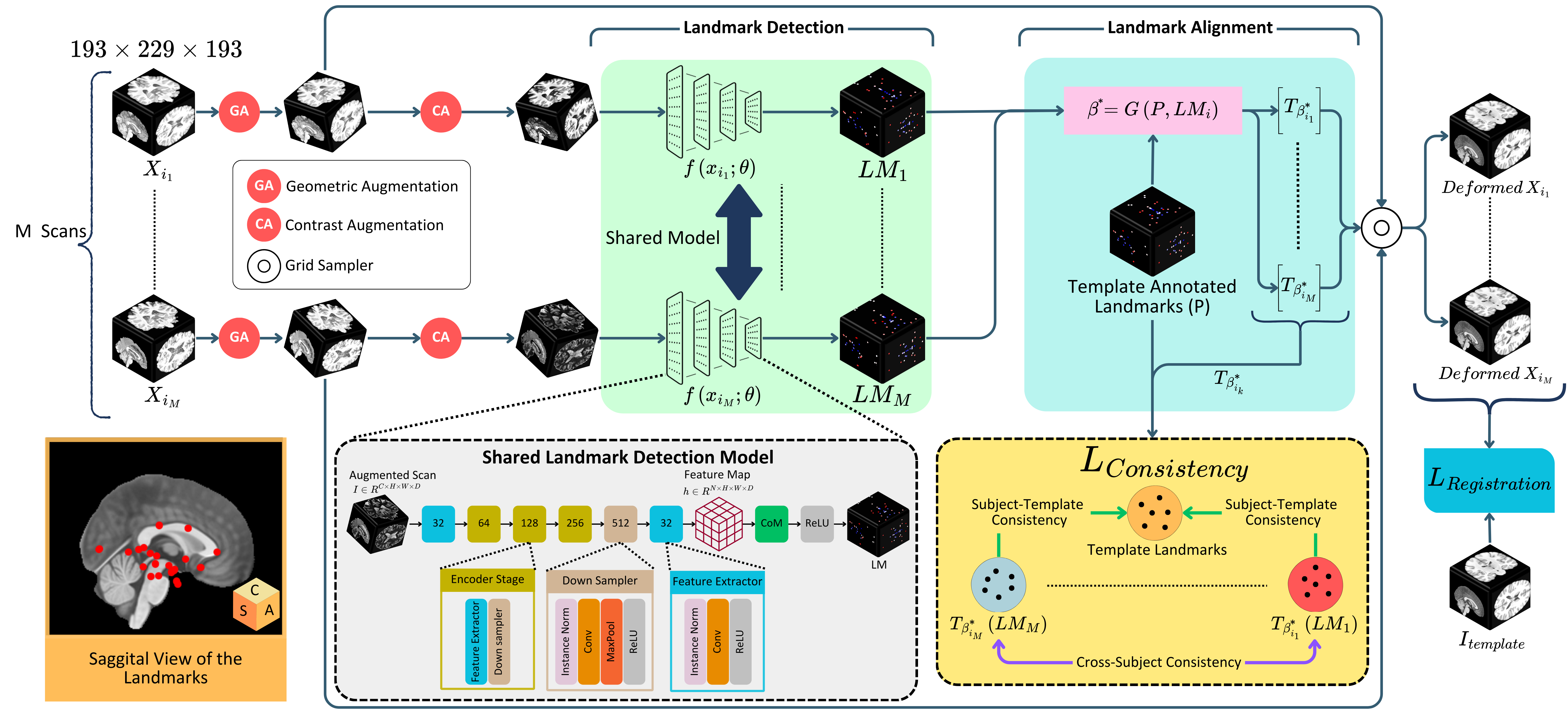}
\caption{Overview of the proposed framework and the consistency-based learning approach. For simplicity, 3D scans are presented in a 2D perspective for visualization.}
\label{SampleOutput}
\end{figure*}

Figure. \ref{SampleOutput} provides an overview of our proposed framework. We begin by randomly selecting \( M \) volumetric scans from the training set \( \boldsymbol{{X}}\). These scans undergo contrast and geometric augmentations before being processed by the landmark detection model \( \boldsymbol{f}(\cdot; \theta) \), which predicts the anatomical landmarks. For each scan, transformations are then computed based on the template landmarks \( \textbf{\textit{P}} \) and the predicted landmarks, aligning the volumetric scan and predicted landmarks to the template space. Finally, two distinct loss functions are calculated to train the model effectively.

\subsection{Training Loss}

Let \( \boldsymbol{X_M} = \{x_{i_1}, x_{i_2}, \ldots, x_{i_M}\} \), where \( \{i_1, i_2, \ldots, i_M\} \subset \{1, \ldots, K\} \), be $M$ random samples from the whole $K$ samples in the training set. Also, $T_{\beta}$ denotes the analytical and differentiable deformation field with control parameters $\beta$, as presented in the \textit{Supplementary Materials}.

For each of the images \( x_{i} \in X_M\), we obtain $L$ anotmical landmarks:  \( \{\boldsymbol{f}(x_{i_1}; \theta), \boldsymbol{f}(x_{i_2}; \theta), \ldots, \boldsymbol{f}(x_{i_M}; \theta)\} \). Subsequently, $M$ deformation fields were obtained: \(\{T_{\beta_{i_1}^*}, \ldots, T_{\beta_{i_M}^*}\} \), where $\beta^* = G\left(\boldsymbol{P}, \boldsymbol{f}(x_{i}; \theta)\right)$ and \( G \) is the function to calculates the parameter \( \beta^* \) based on the pairs of template  and predicted landmarks (\( \boldsymbol{P} \) vs. \( \boldsymbol{f}(x_i; \theta) \)) and the pre-determined spatial transformation model. The interpolation requirements can be relaxed (e.g., to handle noise) by incorporating a regularization term $\lambda > 0$:

\begin{equation}
\label{DefFieldEq}
    G(\boldsymbol{P}, \boldsymbol{f}(x_i; \theta)) =\arg \min_{\beta} \sum_{j=1}^{L} (T_{\beta}(\mathbf{\boldsymbol{f}(x_i; \theta)}^{(j)}) - \boldsymbol{P}^{(j)})^2 + \lambda I,
\end{equation}

\noindent where $\lambda$ is a hyperparameter that determines the level of regularization. For the thin-plate splines (TPS) transformation model, when $\lambda \rightarrow \infty$, the optimal transformation tends to be an affine transformation (i.e., zero bending energy). During the training process, this parameter can be drawn from a distribution \( \lambda \sim N(\lambda) \), learning both affine and nonlinear transformations.

\noindent \textbf{Registration Loss:} The registration loss function, $\mathcal{L}_{\text {registration}}$ of our training to align the subject's scan ($\boldsymbol{x}_{i_k}$) to the template image ($I_{\text{template}}$) is defined as:

\begin{equation}
\label{SimLoss}
\mathcal{L}_{\text{registration}} = \frac{1}{M} \sum_{k=1}^{M} \mathcal{L}_{\text{sim}}\left(x_{i_k} \circ T_{\beta_{i_k}^*}, I_{\text{template}}\right),
\end{equation}

\noindent where $\mathcal{L}_{\text{sim}}$ and $\circ$ denote similarity loss function and spatial transformation operation, respectively.

\noindent \textbf{Consistency Loss:} To regularize the landmark predictions across the training samples, we include the cross-subject landmark position consistency loss ${L}_{\text{consistency}_1}$ and the subject-template counterpart ${L}_{\text{consistency}_2}$:

\begin{equation}
\begin{split}
\mathcal{L}_{\text{consistency}_1} = \frac{1}{\binom{M}{2}} \sum_{1 \leq r < j \leq M} 
\left\| T_{\beta_{i_r}^*}(\boldsymbol{f}(x_{i_r}; \theta)) \right. \\ 
- \left. T_{\beta_{i_j}^*}(\boldsymbol{f}(x_{i_j}; \theta)) \right\|_2,
\end{split}
\end{equation}

\begin{equation}
\mathcal{L}_{\text{consistency}_2} = \frac{1}{M} \sum_{k=1}^{M} \left\| T_{\beta_{i_k}^*}(\boldsymbol{f}(x_{i_k}; \theta)) - \textbf{P} \right\|_2,
\end{equation}

\begin{equation}
\mathcal{L}_{\text{consistency}} = \mathcal{L}_{\text{consistency}_1} + \mathcal{L}_{\text{consistency}_2},
\end{equation}

\noindent
As clinical anatomical landmarks typically follow a pre-defined protocol, this requires inter-subject consistency in point placement regardless of individual anatomical variations. We hypothesize that \( \mathcal{L}_{\text{consistency}_1}\) should reinforce this requirement while \(\mathcal{L}_{\text{consistency}_2} \) ensures the consistency of the predicted landmark protocol with the reference template.

\noindent \textbf{Total Training Loss:} The total training loss for our framework is defined as:

\begin{equation}
\label{tot:loss}
\mathcal{L}_{\text{total}} = (1-\alpha)\mathcal{L}_{\text{registration}} + \alpha\mathcal{L}_{\text{consistency}},
\end{equation}

\noindent
In multi-task learning, sub-task interference is a common challenge \cite{yu2020gradient}. In our case, there is an observed interference between the landmark prediction and registration tasks. To address this issue, we build on insights from prior research \cite{omidi2024unsupervised2,omidi2024unsupervised,sharifzadeh2024mitigating,wang2021survey} and introduce the parameter \(\alpha\) to balance objectives and guide the optimizer more effectively. We define \(\alpha\) as follows:

\begin{equation} 
\label{alpha}
\alpha = \frac{2}{1 + \exp(-5 \eta)} - 1, 
\end{equation}

\noindent where

\begin{equation} 
\eta  = \frac{\textit{current training iteration}}{\textit{total training iterations}}, 
\end{equation}

\noindent
During training, the growing \(\alpha\) is linked to the ratio \(\eta \), which tracks the progress of training iterations. This allows the optimizer to initially emphasize $\mathcal{L}_{\text {registration }}$ for the registration task to gain an understanding of anatomical features and then shift the focus toward refining anatomical landmark predictions. This approach resembles curriculum learning \cite{wang2021survey,bengio2009curriculum} and reduces convergence risk to local minima in the early epochs, promoting more accurate and robust learning.

\subsection{Anatomical Landmark Detection Model}
\label{LandmarkModel}

We utilized a 3D CNN for landmark detection, consisting of nine convolutional blocks, each incorporating 3D convolutional layers, instance normalization, ReLU activations, and max pooling for feature downsampling. The filter count starts at 32 and increases progressively to 512, capturing features at multiple scales. In the final block, the feature channels are reduced to 32, enabling compact yet informative representations. The network processes the input image $I \in \mathbb{R}^{C \times H \times W \times D}$ and produces a feature map $h \in \mathbb{R}^{N \times H \times W \times D}$, where $C$ is the number of input channels, and $H$, $W$, and $D$ represent the height, width, and depth of the scan, respectively, while $N$ is the number of output feature maps. Rather than producing a single output, this network architecture generates multiple feature maps in the final layer, each representing a specific anatomical landmark. These feature maps are then passed through a center of mass (CoM) layer \cite{sofka2017fully,ma2020volumetric}, which computes the center of mass for each feature map and regressed landmark coordinates (as an ordered list). The full network architecture is depicted at the bottom of the Figure. \ref{SampleOutput}.

\subsection{Contrast and Geometric Augmentation}
To develop a contrast-agnostic registration framework, we introduce a method that emulates a wide range of image contrasts using random convolution (RC) \cite{xu2021robust} for contrast augmentation in 3D radiological tasks for the first time. RC has been beneficial in creating robust representations for domain generalization across both computer vision and medical imaging tasks in 2D, such as segmentation and classification \cite{sideri2023mad,wang2024car,ouyang2022causality}. This technique utilizes randomly initialized convolutional filters to modify the visual characteristics of input images while preserving their geometric and structural integrity. Previous research indicates that using large kernels in RC can lead to blurring, which can be problematic for registration tasks that depend on detailed image features to establish pixel-level correspondences. To address this issue, we employ a $1 \times 1 \times 1$ kernel size to minimize artifacts and retain structural details. To cover a broad range of contrast variations, we sequence a set of RC layers with LeakyReLU activation, enabling the network to model complex, non-linear intensity relationships among various MRI contrasts. The convolutional kernel parameters in each RC layer are independently drawn from uniform distributions, ensuring a broad range of contrast transformations. RC enables the network to learn landmarks in a contrast-invariant manner, helping to improve the generalization and stabilize the training process. Samples of contrast-augmented scans with RC are shown in the \textit{Supplementary Materials}. Moreover, we performed random affine augmentations on the scans as our geometric augmentation during training. This augmentation ensures that the model learns to detect landmarks accurately, even under varying spatial configurations, improving generalization to unseen data.

\section{Experiments}
\label{sec:experiments}

\subsection{Data and Pre-Processing}
\label{sec:Datasets}
\noindent \textbf{Training data:} For the reference template (\( I_{\text{template}}\)), we used the widely adopted T1w ICBM152-MNI2009c Symmetric brain MRI template \cite{fonov2009unbiased}, along with 32 expert-annotated landmarks ($\boldsymbol{P}$) based on the AFIDs protocol \cite{taha2023magnetic}. Although other brain landmark protocols also exist, we have chosen AFIDs due to its clinical significance and being the only public dataset of this kind. To train our model, we have curated multi-center datasets of balanced sexes and ages (18-90 yo) to build our training set. For our training set, we obtained 1544 brain MRI datasets (3T \& 7T) by combining 581 T1w brain MRIs from the IXI dataset\footnote{https://brain-development.org/ixi-dataset}, 105 7T T1w brain MRIs from the AHEAD dataset \cite{alkemade2020amsterdam}, 634 T1w brain MRIs from the HCP-A dataset \cite{van2013wu}, and 224 T1w brain MRIs from OpenNeuro \cite{setton2023age2, spreng2022neurocognitive}. We conduct standard pre-processing for all scans: resizing and resampling to $193 \times 229 \times 193$ voxels and $1 \times 1 \times 1$ $mm^{3}$ resolution, rescaling intensity values to [0, 1], skull-stripping using BEaST \cite{eskildsen2012beast}, applying an N4 bias field correction, and affine registration to the ICBM152 space \cite{grabner2006symmetric}. 

\noindent \textbf{Testing data:} To evaluate the effectiveness of our method, we have utilized 122 scans from 4 different sources: 1) 30 T1w MRI scans from the HCP dataset \cite{van2013wu} acquired on a 3T scanner, 2) 30 T1w MRI scans from the publicly available Open Access Series of Imaging Studies (OASIS-1) \cite{marcus2007open} database acquired on a 3T scanner, 3) 32 T1w MRI scans from the Stereotactic Neurosurgery (SNSX) dataset \cite{taha2023magnetic} acquired on a 7T scanner, and 4) 30 T2w MRI scans from the HCP dataset. For each scan, 32 AFIDs landmarks \cite{taha2023magnetic} were manually labeled by experts, consistent with the template we utilized in training.

\subsection{Implementation}
\label{Implementation}

All learning-based methods were implemented using PyTorch. We utilized the Adam optimizer with an initial learning rate of \(10^{-4} \) and a cosine annealing scheduler with a minimum learning rate of \( 10^{-6} \). To balance computational efficiency, \( M \) was set to 2. The model was trained for 2500 epochs. Also, we used TPS as our deformation field ($T_{\beta}$ ) model and mean-squared error (MSE) as a similarity loss function (\(\mathcal{L}_{\text{sim}}\)).
For TPS, the regularization parameter ($\lambda$) was randomly sampled from a log-uniform distribution ranging between 0 and 10. For random affine augmentation, the parameters are uniformly sampled as follows: rotations within [$-180^\circ$, $+180^\circ$], translations between [$-15$, $15$] voxels, scaling factors in [$0.8$, $1.2$], and shear values in the interval [$-0.1$, $0.1$]. Lastly, the RC-based contrast augmentation includes five RC layers, each with kernel weights drawn from a random uniform distribution \( U(0, 2) \). After sampling, these kernel weights are adjusted to be zero-centered. Each RC layer is followed by a LeakyReLU activation function with a negative slope of 0.2.

\subsection{Evaluation Metrics and Baselines}
To assess the effectiveness of our framework, we employed mean radial error (MRE) and success detection rate (SDR). MRE calculates the average Euclidean distance between predictions (\( y \)) and ground truths (\( \hat{y} \)). 
SDR measures the percentage of predictions within a predefined threshold \( \tau  \) from the ground truths, given by \( SDR_{\tau} = \frac{1}{S \times L} \sum_{j=1}^S \sum_{i=1}^L \mathbf{1} \left( \| y_j^i - \hat{y}_j^i \|_2 < \tau \right) \times 100 \% \), where the indicator function \( \mathbf{1}(\cdot) \) returns 1 if the Euclidean distance between \( y \) and \( \hat{y} \) is less than \( \tau \), and 0 otherwise and $S$ and $L$ are representing the number of test scans and number of landmarks in each scan, respectively. For these metrics, lower MRE values indicate higher positional accuracy, while higher SDR values reflect a greater proportion of predictions within the acceptable threshold, both signaling improved model performance. Furthermore, we evaluated statistical significance using paired t-tests to compare the performance of our method against baseline methods. A p-value less than 0.05 indicates statistical significance.

We compare our method against three classic SOTA traditional methods (ANTs \cite{avants2009advanced}, NiftyReg \cite{modat2010fast}, and 3D SIFT \cite{rister2017volumetric}), four SOTA learning-based approaches (KeyMorph \cite{wang2023robust}, BrainMorph \cite{wang2024brainmorph}, uniGradICON \cite{tian2024unigradicon}, MultiGradICON \cite{demir2024multigradicon}), and one fully supervised CNN. For ANTs, we implemented two versions: one using the MI metric and the other using the cross-correlation (CC) metric, both with a multi-resolution approach (10x50x50x20) and SyN (Symmetric Normalization) transformation with a gradient step size of 0.25. For NiftyReg, we employed its non-rigid registration algorithm with the normalized mutual information (NMI) metric. In all registration algorithms, the deformation fields from template-to-subject registration were applied to warp the landmarks from the template space to the subject space for prediction calculation. For 3D SIFT, we used SIFT features in the template space to locate corresponding landmarks in the subject space by identifying the closest match with cosine similarity across 3D SIFT points. For KeyMorph \cite{wang2023robust}, BrainMorph \cite{wang2024brainmorph}, uniGradICON \cite{tian2024unigradicon}, and MultiGradICON \cite{demir2024multigradicon}, we use their official implementations and adopt the optimal hyperparameters as specified in their respective publications. Finally, for the fully supervised CNN, we train a 3D CNN model with the same architecture as our proposed anatomical landmark detection model (detailed in Sec. \ref{LandmarkModel}) in a supervised manner using the labeled datasets described in Sec. \ref{sec:Datasets}. For this model, we use the MSE loss function and maintain the same training parameters as our original training detailed in Sec. \ref{Implementation}. It should be noted that we did not test this supervised CNN on the HCP-T2W dataset because, despite incorporating contrast augmentation during training, the number of annotated T2W samples available was very limited.

\subsection{Results}
Table \ref{tableT1} shows the results across three datasets: HCP T1w, OASIS, and SNSX. The evaluation metrics include MRE in millimeters and SDR (\%) at thresholds of 3mm, 6mm, and 9mm. Our proposed method, \textbf{CABLD}, consistently achieved the best results on all datasets. On HCP T1w, it obtained the lowest MRE of 3.27 ± 2.24 mm and the highest SDRs of 54.48\% at 3 mm, 93.69\% at 6 mm, and 98.94\% at 9 mm. For the OASIS dataset, \textbf{CABLD} recorded an MRE of 3.89 ± 2.69 mm with SDRs of 39.24\%, 87.15\%, and 98.22\%. On SNSX, the method achieved an MRE of 5.11 ± 3.19 mm and SDRs of 29.63\%, 71.01\%, and 91.44\%. Compared to other methods, \textbf{CABLD} outperformed both traditional techniques, like 3D SIFT and NiftyReg, and DL-based methods, such as KeyMorph, BrainMorph, uniGradICON, and MultiGradICON ($p<0.05$). Importantly, \textbf{CABLD} could outperform the supervised 3D CNN with statistical significance ($p<0.05$).

\begin{table*}[tb]
    \centering
    \caption{Comparing landmark detection test performance across three datasets: HCP T1w, OASIS, and SNSX.}
    \resizebox{\textwidth}{!}{%
    \begin{tabular}{cc|cccc|cccc|cccc}
        \toprule
        \multirow{2}{*}{\textbf{Method}} & \multirow{2}{*}{\textbf{Optimization Metric}} 
        & \multicolumn{4}{c|}{\textbf{HCP T1w}} 
        & \multicolumn{4}{c|}{\textbf{OASIS}} 
        & \multicolumn{4}{c}{\textbf{SNSX}} \\
        \cline{3-14}
        & 
        & \textbf{MRE (mm) \textcolor{red}{↓}} & \textbf{SDR (3mm) \textcolor{red}{↑}} & \textbf{SDR (6mm) \textcolor{red}{↑}} & \textbf{SDR (9mm) \textcolor{red}{↑}} 
        & \textbf{MRE (mm) \textcolor{red}{↓}} & \textbf{SDR (3mm) \textcolor{red}{↑}} & \textbf{SDR (6mm) \textcolor{red}{↑}} & \textbf{SDR (9mm) \textcolor{red}{↑}} 
        & \textbf{MRE (mm) \textcolor{red}{↓}} & \textbf{SDR (3mm) \textcolor{red}{↑}} & \textbf{SDR (6mm) \textcolor{red}{↑}} & \textbf{SDR (9mm) \textcolor{red}{↑}} \\
        \hline
        3D SIFT \cite{rister2017volumetric} & - 
        & 39.44 ± 31.02 & 5.72\% & 20.62\% & 26.97\%
        & 39.08 ± 29.70 & 3.70\% & 17.71\% & 24.88\%
        & 41.67 ± 31.84 & 4.52\% & 17.13\% & 25.43\% \\
        NiftyReg \cite{modat2010fast} & NMI 
        & 4.43 ± 2.42  & 25.00\% & 81.25\% & 96.04\%
        & 8.23 ± 3.29 & 1.85\% & 22.69\% & 65.16\%
        & 9.61 ± 4.03 & 0.43\% & 12.71\% & 50.11\% \\
        ANTs \cite{avants2009advanced}& CC 
        & 3.85 ± 2.26 & 36.97\% & 89.16\% & 97.54\%
        & 4.38 ± 2.64 & 29.39\% & 81.25\% & 97.33\%
        & 6.36 ± 3.28 & 10.88\%  & 49.78\% & 87.39\% \\
        ANTs \cite{avants2009advanced}& MI 
        & 3.65 ± 2.29 & 45.52\% & 92.29\% & 98.85\%
        & 4.15 ± 2.65 & 38.88\% & 85.38\% & 97.91\%
        & 6.06 ± 3.22 & 18.75\% & 54.43\% & 90.73\% \\
        \midrule

        KeyMorph (64 KPs) \cite{wang2023robust}& Dice
        & 8.05 ± 4.51 & 10.93\%  & 38.43\% & 62.60\%
        & 8.20 ± 4.64 & 10.30\% & 36.57\% & 63.88\%
        & 9.73 ± 5.35 & 6.03\% & 31.35\% & 57.86\% \\

        KeyMorph (128 KPs) \cite{wang2023robust}& Dice
        & 5.77 ± 2.91 & 13.95\% & 58.43\% & 89.47\%
        & 6.41 ± 3.41 & 13.31\% & 51.62\% & 81.71\%
        & 8.99 ± 4.16 & 3.66\%  & 25.43\% & 54.95\% \\
        
        KeyMorph (256 KPs) \cite{wang2023robust}& Dice
        & 5.37 ± 3.12 & 20.83\% & 66.04\% & 88.96\%
        & 6.44 ± 3.81 & 12.61\% & 55.20\% & 81.94\%
        & 8.80 ± 5.22 & 7.65\%  & 35.56\% & 59.48\% \\
        
        KeyMorph (512 KPs)  \cite{wang2023robust} & Dice
        & 4.67 ± 2.47 & 23.30\% & 78.12\% & 95.80\%
        & 7.15 ± 3.63 & 6.82\% & 40.97\% & 75.81\%
        & 5.77 ± 3.27 & 18.10\% & 60.66\% & 87.82\% \\

        BrainMorph \cite{wang2024brainmorph} & MSE+Dice
        & 4.11 ± 2.30 & 31.35\%  & 86.15\% & 97.81\%
        & 5.28 ± 3.07 & 17.36\% & 74.31\% & 90.39\%
        & 13.66 ± 18.21 & 14.66\% & 41.38\% & 61.85\% \\
	
        uniGradICON \cite{tian2024unigradicon} & $\text{LNCC}$
        & 4.12 ± 2.53 & 34.38\% & 84.06\% & 97.29\%
        & 4.63 ± 3.00 & 30.09\%  & 76.97\%  & 93.29\% 
        & 5.27 ± 3.53 & 29.53\% & 70.63\% & 88.90\% \\

        MultiGradICON \cite{demir2024multigradicon} & $\text{LNCC}^2$
        & 4.10 ± 2.56 & 34.90\% & 84.17\% & 96.67\%
        & 4.62 ± 3.01 & 30.79\% & 77.89\% & 93.52\%
        & 5.21 ± 3.40 & 28.68\% & 70.84\% & 89.12\%\\

        \midrule

        Fully Supervised 3D CNN & -
        & 4.65 ± 2.40 & 24.27\% & 72.81\% & 95.42\%
        & 4.53 ± 2.81 & 25.00\% & 79.28\% & 94.44\%
        & 6.64 ± 3.86 & 12.61\% & 52.26\% & 76.19\%\\
        
        \midrule
        \rowcolor[rgb]{0.85, 0.92, 0.83}
        \textbf{CABLD (Ours)} & MSE
        & \textbf{3.27 ± 2.24} & \textbf{54.48\%} & \textbf{93.69\%} & \textbf{98.94\%}
        & \textbf{3.89± 2.69} & \textbf{39.24\%} & \textbf{87.15\%} & \textbf{98.22\%}
        & \textbf{5.11 ± 3.19} & \textbf{29.63\%} & \textbf{71.01\%} & \textbf{91.44\%} \\
        \bottomrule
    \end{tabular}%
    }
    \label{tableT1}
\end{table*}

\begin{figure*}[tb]
\centering
\includegraphics[scale=0.275]{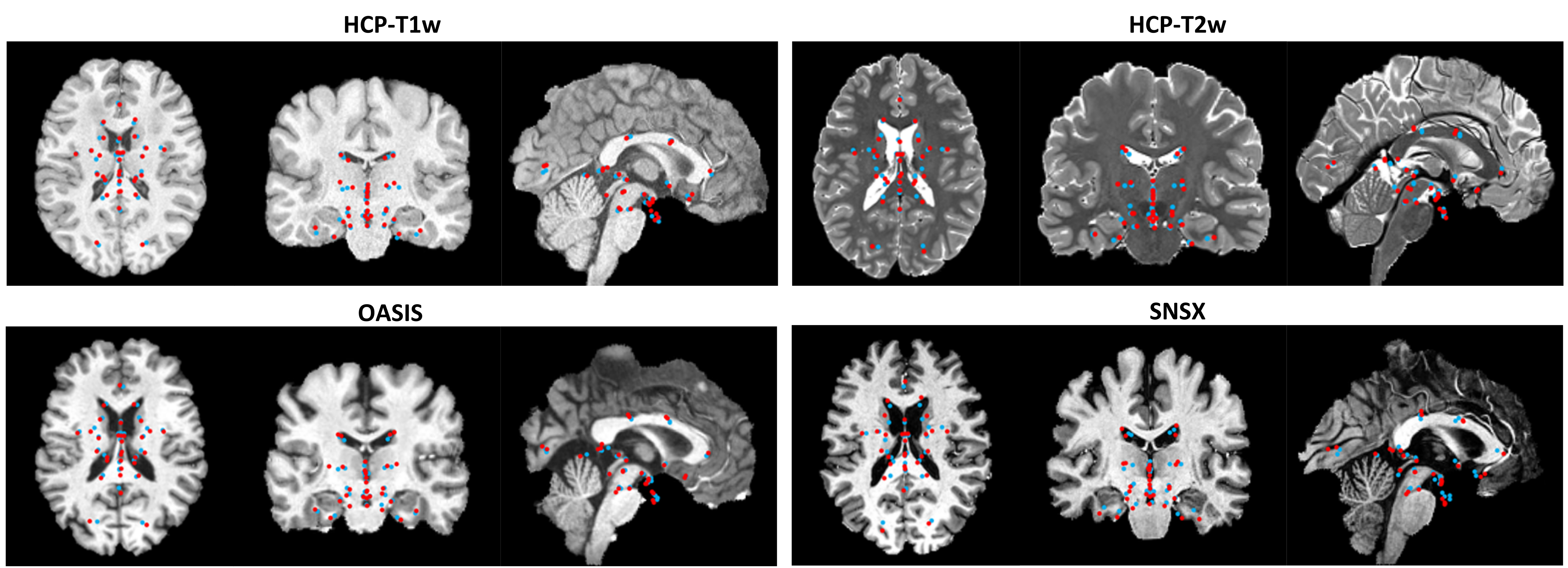}
\caption{Qualitative comparison of anatomical landmark detection results using the proposed technique, illustrated on coronal, axial, and sagittal mid-slices (red=ground truths, blue=automatic results) for samples across all datasets. Note that the landmarks are shown as projections of 3D points in each canonical view for visual demonstration purposes only.}
\label{QualiRes}
\end{figure*}

To check if \textbf{CABLD} can generalize on other MRI contrasts, we evaluated it on the HCP T2w dataset. Table \ref{tableT2} shows the results of different models on the HCP T2w dataset. Although our model has only been trained on T1w scans, we can see that it could outperform the other baseline methods on the T2w scans. Specifically, \textbf{CABLD} achieved an MRE of 3.99 ± 2.25 mm, which is comparable to the best-performing method, ANTs (MI), at 3.91 ± 2.19mm. Notably, \textbf{CABLD} attained the highest SDR at the 6mm (86.43\%) and 9mm (98.99\%) thresholds, surpassing all other methods. While its SDR at the 3 mm threshold (27.19\%) was slightly lower than that of ANTs (MI) (35.00\%) and MultiGradICON (33.33\%), \textbf{CABLD} still demonstrated strong performance, given it was not trained on T2w images. Figure. \ref{QualiRes} illustrates a few sample of the \textbf{CABLD} results.

\begin{table*}[tb]
    \centering
    
    \scriptsize


    \caption{Comparing landmark detection performance on HCP-T2w test dataset.}
    \begin{tabular}{cc|cccc}
        \toprule
        \textbf{Method} & \textbf{Optimization Metric} & \textbf{MRE (mm) \textcolor{red}{↓}} & \textbf{SDR (3mm) \textcolor{red}{↑}} & \textbf{SDR (6mm) \textcolor{red}{↑}} & \textbf{SDR (9mm) \textcolor{red}{↑}} \\
        \midrule
        3D SIFT \cite{rister2017volumetric} & - & 54.90 ± 24.51  & 0.00\% & 0.73\% & 1.35\% \\
        NiftyReg \cite{modat2010fast}&  NMI & 4.40 ± 2.41 & 25.50\%  & 82.25\% & 96.28\%\\
        ANTs \cite{avants2009advanced} & MI & \textbf{3.91 ± 2.19}  & \textbf{35.00\%} & 84.27\% & 98.64\% \\
        \midrule
        KeyMorph (64 KPs) \cite{wang2023robust}& Dice & 6.00 ± 2.64  & 11.87\%  & 52.08\% & 88.33\% \\
        KeyMorph (128 KPs) \cite{wang2023robust} &  Dice & 8.66 ± 4.29  & 5.93\% & 28.64\% & 58.23\% \\
        KeyMorph (256 KPs) \cite{wang2023robust} &  Dice & 6.41 ± 3.06  & 8.65\% & 51.56\% & 83.54\% \\
        KeyMorph (512 KPs) \cite{wang2023robust}& Dice & 5.54 ± 3.31 & 22.18\% & 64.06\% & 85.62\% \\

        BrainMorph \cite{wang2024brainmorph} & MSE+Dice
        & 4.24 ± 2.43 & 32.71\% & 82.19\% & 96.77\% \\
	
        uniGradICON \cite{tian2024unigradicon} & $\text{LNCC}$
        & 13.44 ± 3.88 & 0.42\% & 3.33\% & 12.60\% \\

        MultiGradICON \cite{demir2024multigradicon} & $\text{LNCC}^2$
        & 4.31 ± 2.70 & 33.33\% & 81.83\% & 95.83\% \\



         \midrule
         \rowcolor[rgb]{0.85, 0.92, 0.83}
        \textbf{CABLD (Ours)} & MSE & 3.99 ± 2.25 &  27.19\% & \textbf{86.43\%} & \textbf{98.99\%} \\
        \bottomrule
    \end{tabular}
    \label{tableT2}
\end{table*}

\subsection{Ablation Studies}
\noindent \textbf{Influence of RC and $\mathcal{L}_{\text{consistency}}$:} We conducted ablation studies to assess the impact of the consistency loss ($\mathcal{L}_{\text{consistency}}$) and contrast augmentation with RC on our model's performance, as shown in Table \ref{tab:influence_mre_datasets}. Adding $\mathcal{L}_{\text{consistency}}$ to the base model drastically reduced the Mean Radial Error across all datasets. For example, on HCP T1w, the MRE decreased from 53.69 ± 25.63 mm to 3.70 ± 2.41 mm. Incorporating RC further improved the MRE, notably reducing it to 3.99 ± 2.25 mm on HCP-T2w. These results demonstrate that both $\mathcal{L}_{\text{consistency}}$ and RC are essential for enhancing the model's accuracy and generalization across different datasets. During our exploration, we observed that the two consistency terms ($\mathcal{L}_{\text{consistency}_1}$ and $\mathcal{L}_{\text{consistency}_2}$) form a triangular relationship among three scans, and the model training failed to converge when only one of these sub-losses was used. Consequently, we did not conduct separate ablation studies for them.

\begin{table}[tb]
    \centering
    \caption{Testing the influence of RC and consistency loss 
    ($\mathcal{L}_{\text{consistency}}$) on the proposed framework 
    using landmark MRE (mm) across all datasets.}
    \renewcommand{\arraystretch}{1.2}
    \setlength{\tabcolsep}{4pt}     

    \resizebox{\columnwidth}{!}{
    \begin{tabular}{lcccc}
        \toprule
        \textbf{Methods} & \textbf{HCP-T1w} & \textbf{OASIS} & 
        \textbf{SNSX} & \textbf{HCP-T2w} \\ 
        \midrule
        Base Model & 
        53.69 ± 25.63 & 55.02 ± 25.68 & 53.11 ± 29.18 & 59.79 ± 26.39 \\
        \midrule
        +$\mathcal{L}_{\text{consistency}}$ & 
        \makecell{3.70 ± 2.41\\\textcolor{blue}{(-49.99)}} & 
        \makecell{4.03 ± 2.69\\\textcolor{blue}{(-50.99)}} & 
        \makecell{6.43 ± 3.64\\\textcolor{blue}{(-46.68)}} & 
        \makecell{45.90 ± 17.77\\\textcolor{blue}{(-13.89)}} \\
        \midrule
        +RC & 
        \makecell{3.27 ± 2.24\\\textcolor{blue}{(-0.43)}} & 
        \makecell{3.89 ± 2.69\\\textcolor{blue}{(-0.14)}} & 
        \makecell{5.11 ± 3.59\\\textcolor{blue}{(-1.32)}} & 
        \makecell{3.99 ± 2.25\\\textcolor{blue}{(-41.91)}} \\ 
        \bottomrule
    \end{tabular}
    } 
    \label{tab:influence_mre_datasets}
\end{table}

\begin{figure}[ht!]
\centering
\includegraphics[scale=0.33]{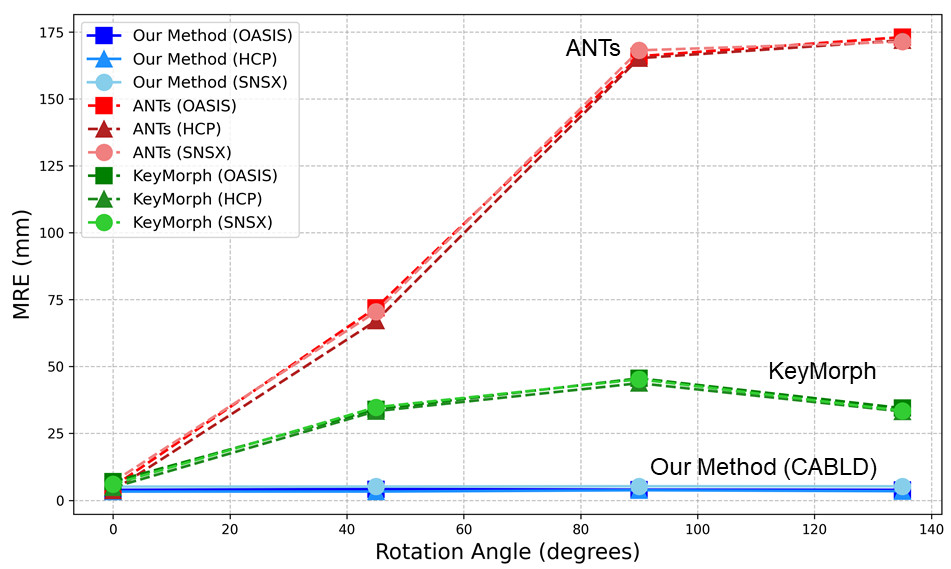}
\caption{Comparison of different methods' MRE (mm) performance across varying levels of added rotation misalignments.}
\label{PerfComp}
\end{figure}

\noindent \textbf{Robustness against rotation misalignments:} Different head orientations in the MRI scans can occur due to patient positioning, between-scan movement, and scanner setup, but they could affect the quality of landmark detection \cite{billot2024se,singh2024data}. Therefore, to assess the robustness of \textbf{CABLD} against this factor, we have evaluated its performance under conditions of augmented rotation misalignments to each scan. Figure \ref{PerfComp} illustrates the MRE as a function of the augmented rotation. Our analysis reveals that while methods such as ANTs perform adequately with very minimal augmented misalignment, their performance declines quickly as the additional rotation increases. In contrast, \textbf{CABLD} demonstrates robust and consistent performance across a wide range of transformations, even under large misalignments, without notable performance drops.

\noindent \textbf{Sensitivity to template choice:} In our experiments, we used ICBM152, the most widely adopted brain MRI atlas, derived from 152 young adults \cite{grabner2006symmetric}. To assess its impact, we analyzed the results from the complete test set, categorizing the subjects into three age groups (20–40, 40–60, 60–90 years). The MREs (mm) were 3.65 ± 1.56, 4.38 ± 2.00, and 4.38 ± 0.81 for the age groups, respectively. An ANOVA test indicated \textbf{no significant accuracy differences between the age groups} ($p = 0.07 > 0.05$), confirming our framework’s robustness despite using the ICBM152 atlas. In addition, we also validated our method's robustness when using the T1w Colin27 atlas as the template, with the results in the \textit{Supplementary Materials}.

\subsection{Downstream tasks using detected landmarks}
Abbas et al. \cite{abbass2022application} demonstrated that AFIDs-based inter-landmark distances from expert labels can serve as potential biomarkers for Parkinson's disease (PD). Following a similar approach, we generated landmarks using our model for 30 healthy subjects from the OASIS dataset and 30 PD subjects from PPMI \cite{marek2011parkinson}. To confirm the effectiveness of inter-landmark distances from our DL method as PD biomarkers, we conducted Wilcoxon tests with Bonferroni correction, identifying \textbf{117} out of 496 distances as significantly different between healthy vs. PD cohorts ($p < 0.05/496$). Furthermore, to assess the diagnostic utility of the generated landmarks for PD detection, we applied our model to 210 MRIs from the PPMI dataset \cite{marcus2007open} (148 PD vs. 62 healthy). Using the predicted inter-landmark distances, we trained a support vector machine (SVM) classifier, achieving a good F1-score of \textbf{81.2 ± 8.8\%} in PD detection via a 10-fold cross-validation. Both experiments confirmed the values of the landmarks generated by our model for PD diagnosis.

Inspired by prior studies that used brain landmarks for Alzheimer’s disease (AD) detection \cite{zhang2017alzheimer,liu2018landmark,liu2020anatomical,liu2018anatomical,zhang2017landmark}, we performed a study to assess the effectiveness of our model-generated landmarks for the same task. We predicted landmarks for MRIs of 52 AD patients from the ADNI dataset \cite{jack2008alzheimer} and 62 healthy subjects in the same age range from the PPMI dataset. Using the inter-landmark distances, we trained an SVM classifier, achieving a F1-score of \textbf{93.5 ± 6.3\%} via a 10-fold cross-validation. Our findings validate the potential of our technique for accurate AD diagnosis.

\section{Discussion}
\label{sec:discussion}

Different from the previous landmark-based registration methods \cite{evan2022keymorph,wang2024brainmorph,wang2023robust,bhalodia2020self,bhalodia2021leveraging}, which produce spontaneous anatomical landmarks to explain registration results and regularize image alignment, \textbf{CABLD} specifically focuses on accurate anatomical landmarks that are consistent with a pre-defined protocol. We achieve this by incorporating subject-template and cross-subject consistency loss functions, with the injection of expert knowledge using a single population-averaged MRI atlas. This greatly reduces the demand for large amounts of annotated data in model training. Note that as a reference atlas is involved during the training, which allows landmark discovery in individuals to adapt to the references, our approach deviates from the conventional one-shot learning strategy. We employed an adaptive mixed loss function, gradually shifting the focus from registration to consistency regularization during training. By first learning the general spatial relationship of the anatomical context and then refining predictions to align with anatomically significant landmarks, our framework balances the need for both alignment and anatomical consistency. As our aim is to detect clinically significant anatomical landmarks, \textbf{CABLD} prioritizes accurate landmark correspondence over full volumetric registration. Thus, the resulting deformation fields from the algorithm may not achieve registration accuracy globally, in comparison to registration-specific DL algorithms. Our use of random convolution offers a simpler and more robust way to allow contrast-agnostic landmark detection. Compared with stochastic intensity transformation \cite{mok2024modality}, RC simulates richer and more nuanced contrasts. To the best of our knowledge, we are the first to adopt 3D RC in medical image analysis. In our framework, $M$ training samples can be used to enforce the inter-subject/template consistency and are constrained by the trade-off with the hardware resource. With high-resolution 3D MRIs, we used $M$=2, but a larger $M$ can be adopted for smaller data or better hardware. We will study the impacts in the future.

\section{Conclusion}

We proposed \textbf{CABLD}, a novel framework for contrast-agnostic brain landmark detection that only requires a single reference annotation. Our method leverages consistency-based regularization and RC to accurately detect anatomical landmarks and encourage the generalization to unseen contrasts. Thorough validation has confirmed that our method achieves SOTA accuracy compared with existing methods. Notably, \textbf{CABLD} shows strong robustness against anatomical misalignment from the reference scan.

\noindent \textbf{Acknowledgment} 

\noindent
We acknowledge the support of the Natural Sciences and Engineering Research Council of Canada (NSERC). S.S. is supported by the doctoral training scholarship from the Fonds de recherche du Québec – Santé (\href{https://doi.org/10.69777/373847}{https://doi.org/10.69777/373847}). H.R. is supported by the Fond de la Recherche du Québec  (\href{https://doi.org/10.69777/361263}{https://doi.org/10.69777/361263}). Y.X. is supported by the Fond de la Recherche du Québec – Santé (FRQS-chercheur boursier Junior 1) and Parkinson Quebec (\href{https://doi.org/10.69777/330745}{https://doi.org/10.69777/330745}). We thank Mr. Simon Joseph Clément Crête for his contribution to data curation for this study.

This study has received approval from the Research Ethics Board (REB) of Concordia University.

Data used in the preparation of this article was obtained on 2024-03-21 from the Parkinson’s Progressive Markers Initiative (PPMI) database, RRID: SCR\_006431. For up-to-date information on the study, visit www.ppmi-info.org. 

PPMI – a public-private partnership – is funded by the Michael J. Fox Foundation for Parkinson's Research and funding partners, including 4D Pharma, Abbvie, AcureX, Allergan, Amathus Therapeutics, Aligning Science Across Parkinson’s, AskBio, Avid Radiopharmaceuticals, BIAL, BioArctic, Biogen, Biohaven, BioLegend, BlueRock Therapeutics, Bristol-Myers Squibb, Calico Labs, Capsida Biotherapeutics, Celgene, Cerevel Therapeutics, Coave Therapeutics, DaCapo Brainscience, Denali, Edmond J. Safra Foundation, Eli Lilly, Gain Therapeutics, GE HealthCare, Genentech, GSK, Golub Capital, Handl Therapeutics, Insitro, Jazz Pharmaceuticals, Johnson \& Johnson Innovative Medicine, Lundbeck, Merck, Meso Scale Discovery, Mission Therapeutics, Neurocrine Biosciences, Neuron23, Neuropore, Pfizer, Piramal, Prevail Therapeutics, Roche, Sanofi, Servier, Sun Pharma Advanced Research Company, Takeda, Teva, UCB, Vanqua Bio, Verily, Voyager Therapeutics, the Weston Family Foundation and Yumanity Therapeutics.

The Alzheimer's disease MRI data in this study were obtained from the Alzheimer’s Disease Neuroimaging Initiative (ADNI) database, which is available to all researchers.

\label{sec:conclusion}

{
    \small
    \bibliographystyle{ieeenat_fullname}
    \bibliography{main}
}

\clearpage

\setcounter{page}{1}
\setcounter{section}{0}
\setcounter{equation}{0}
\renewcommand{\thetable}{S\arabic{table}}
\renewcommand{\thefigure}{S\arabic{figure}}
\renewcommand{\theequation}{S\arabic{equation}}
\setcounter{table}{0}
\setcounter{figure}{0}

\maketitlesupplementary

\section{Analytical and Differentiable Coordinate Transformations}
         
\textbf{Notation:} Lowercase bold letters denote column vectors, while uppercase bold letters are used for matrices. Coordinates in $D$ dimensions are represented as column vectors, i.e., $\mathbf{x} \in \mathbb{R}^D$. The symbol $\tilde{\mathbf{x}}$ denotes $\mathbf{x}$ in homogeneous coordinates, expressed as $\tilde{\mathbf{x}} = [\mathbf{x}, 1]^T$. Superscripts like $\mathbf{x}^{(j)}$ are used to indicate distinct instances of $\mathbf{x}$ (such as in a dataset), while subscripts, $\mathbf{x}_j$, represent the $j$-th component of $\mathbf{x}$.

We introduce families of parametric transformations that can be derived explicitly in closed-form based on corresponding landmark pairs. Let us consider $N$ matching landmark pairs $\{(\mathbf{x}^{(j)}, \mathbf{y}^{(j)})\}_{j=1}^{N}$, where $\mathbf{x}^{(j)}, \mathbf{y}^{(j)} \in \mathbb{R}^D$ and $N > D$. For simplicity, we define $\mathbf{X} := \langle \mathbf{x}^{(1)} \ldots \mathbf{x}^{(N)} \rangle \in \mathbb{R}^{D \times N}$, and similarly for $\tilde{\mathbf{X}}$ and $\mathbf{Y}$. We define a transformation function $T_\beta: \mathbb{R}^D \rightarrow \mathbb{R}^D$, where $\beta \in \mathcal{B}$ are the transformation parameters.

\subsection{Thin-Plate Spline Deformation Model}

The thin-plate spline (TPS) transformation is used for coordinate mapping, delivering a non-rigid, parameterized deformation model with a closed-form solution for interpolating corresponding landmarks \cite{donato2002approximate,bookstein1989principal,rohr2001landmark, zhao2022thin}. This approach offers greater adaptability than affine mappings while inherently encompassing affine transformations as a specific case.

The TPS deformation model $T_\beta: \mathbb{R}^D \rightarrow \mathbb{R}^D$ is expressed as:
\begin{equation}
    T_\beta(\mathbf{x}) = \mathbf{W}^T \tilde{\mathbf{x}} + \sum_{j=1}^{N} \mathbf{v}_j \Phi(\|\mathbf{x}^{(j)} - \mathbf{x}\|^2),
\end{equation}
where $\mathbf{W} \in \mathbb{R}^{D \times (D+1)}$ and $\mathbf{v}_j \in \mathbb{R}^D$ represent the transformation parameters ($\beta$), and $\Phi(r) = r^2 \ln(r)$. Additionally, $\mathbf{V} =\{\mathbf{v}_j\}_{j=1}^N$, making the full parameter set $\beta = \{\mathbf{W}, \mathbf{V}\}$.

\noindent

The transformation $T$ minimizes the bending energy:
\begin{equation}
    I_T = \int_{\mathbb{R}^D} \|\nabla^2 T(\mathbf{x})\|_F^2 \, d\mathbf{x},
\end{equation}
which ensures that $T$ is smooth with square-integrable second derivatives. We impose the interpolation conditions $T(\mathbf{x}^{(j)}) = \mathbf{y}^{(j)}$ for $j = 1, \ldots, N$, and the following constraints to ensure a well-posed solution:
\begin{equation}
    \sum_{j=1}^{N} \mathbf{v}_j = \mathbf{0} \quad \text{and} \quad \sum_{j=1}^{N} \mathbf{v}_j (\mathbf{x}^{(j)})^T = \mathbf{0}.
\end{equation}

\noindent
Based on the mentioned conditions, the linear system below can be considered for $\beta$:
\begin{equation}
    \label{LinSys}
    \begin{bmatrix} 
        \mathbf{M} & \mathbf{R} \\ 
        \mathbf{R}^T & \mathbf{Z} 
    \end{bmatrix} 
    \begin{bmatrix} 
        \mathbf{V} \\ 
        \mathbf{W} 
    \end{bmatrix} 
    = 
    \begin{bmatrix} 
        \mathbf{Y} \\ 
        \mathbf{Z} 
    \end{bmatrix},
\end{equation}

\noindent where $\mathbf{M} \in \mathbb{R}^{N \times N}$ with entries $M_{ij} = \Phi(\|\mathbf{x}^{(i)} - \mathbf{x}^{(j)}\|^2)$, $\mathbf{R} \in \mathbb{R}^{N \times (D+1)}$ where each Row $j$ is $\tilde{\mathbf{x}}^{(j)T}$, $\mathbf{V} \in \mathbb{R}^{N \times D}$ with the $j^{th}$ row being $\mathbf{v}_j^T$, $\mathbf{Y} \in \mathbb{R}^{N \times D}$ with row entries of $\mathbf{y}^{(j)T}$, and $\mathbf{Z}$ is a zero matrix with the proper size.

\noindent
Accordingly, the solution $\beta^*$ is obtained by:
\begin{equation}
    \beta^*
    =
    \begin{bmatrix} 
        \mathbf{V}^* \\ 
        \mathbf{W}^* 
    \end{bmatrix}
    = 
    \begin{bmatrix} 
        \mathbf{M} & \mathbf{R} \\ 
        \mathbf{R}^T & \mathbf{Z} 
    \end{bmatrix}^{-1}
    \begin{bmatrix} 
        \mathbf{Y} \\ 
        \mathbf{Z} 
    \end{bmatrix}.
\end{equation}

\noindent Using the equations above, $\beta^*$ can be formulated as a differentiable function, ensuring integration with gradient-based optimization frameworks.

\noindent
Finally, the general TPS equation can be improved (e.g., to handle noise) by incorporating a regularization term:
\begin{equation}
    \beta^* = \arg \min_{\beta} \sum_{j=1}^{N} \|T_\beta(\mathbf{x}^{(j)}) - \mathbf{y}^{(j)}\|^2 + \lambda I_T,
\end{equation}
where $\lambda$ is a positive hyperparameter that determines the regularization level. As $\lambda \rightarrow \infty$, the optimal transformation $T$ tends to an affine form. This can be achieved by modifying the matrix $\mathbf{M}$ to $\mathbf{M} + \lambda \mathbf{I}$ in the linear system (Eq. \ref{LinSys}). The parameter $\lambda$ could influence the solution $\beta^*$, leading it either toward an affine transformation as $\lambda \to \infty$ or toward a fully nonlinear deformation as $\lambda \to 0$.

\begin{figure*}[tb]
\centering
\includegraphics[scale=0.35]{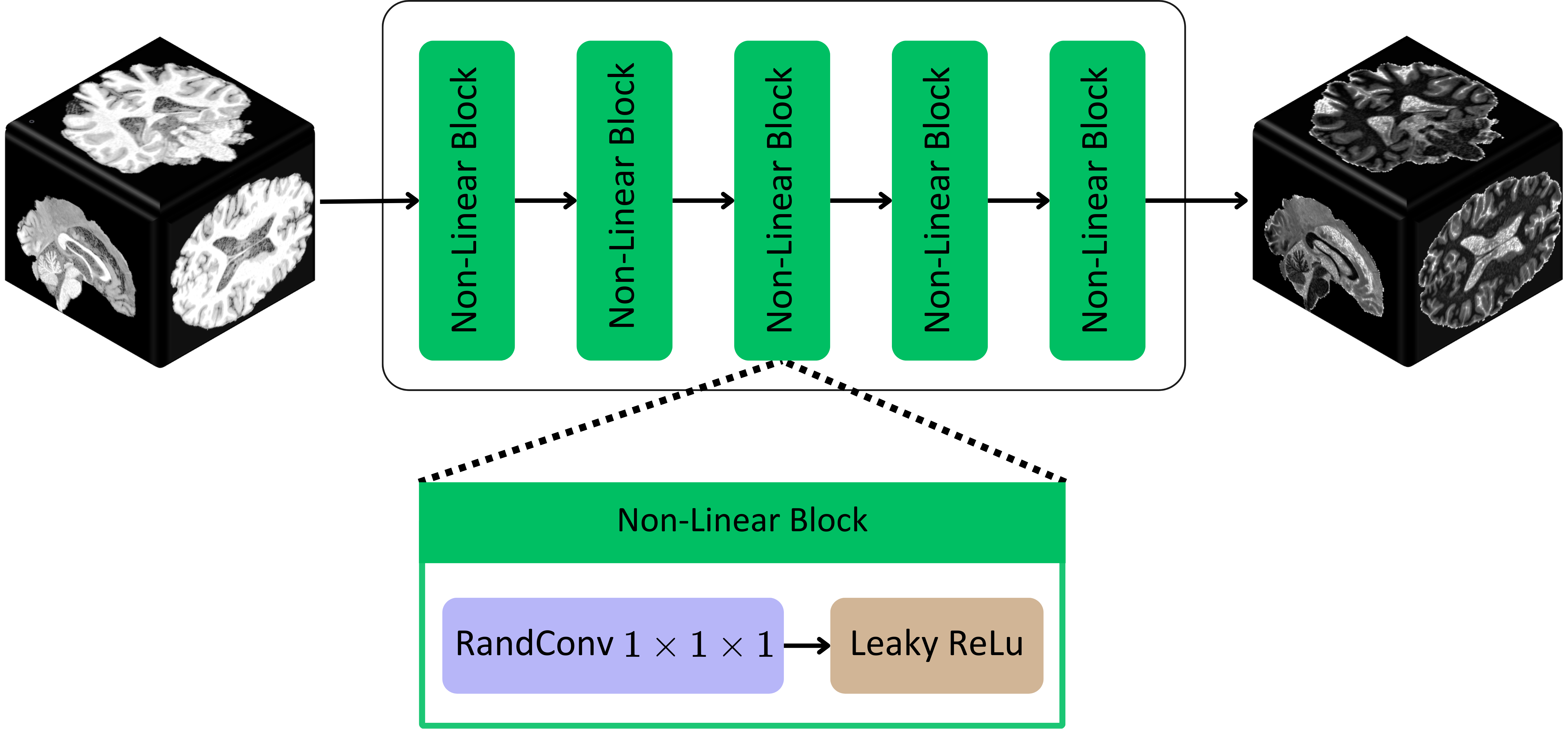}
\caption{The overall architecture of the proposed 3D contrast augmentation method using random convolution.}
\label{RCModel}
\end{figure*}

\begin{figure*}[tb]
\centering
\includegraphics[scale=0.355]{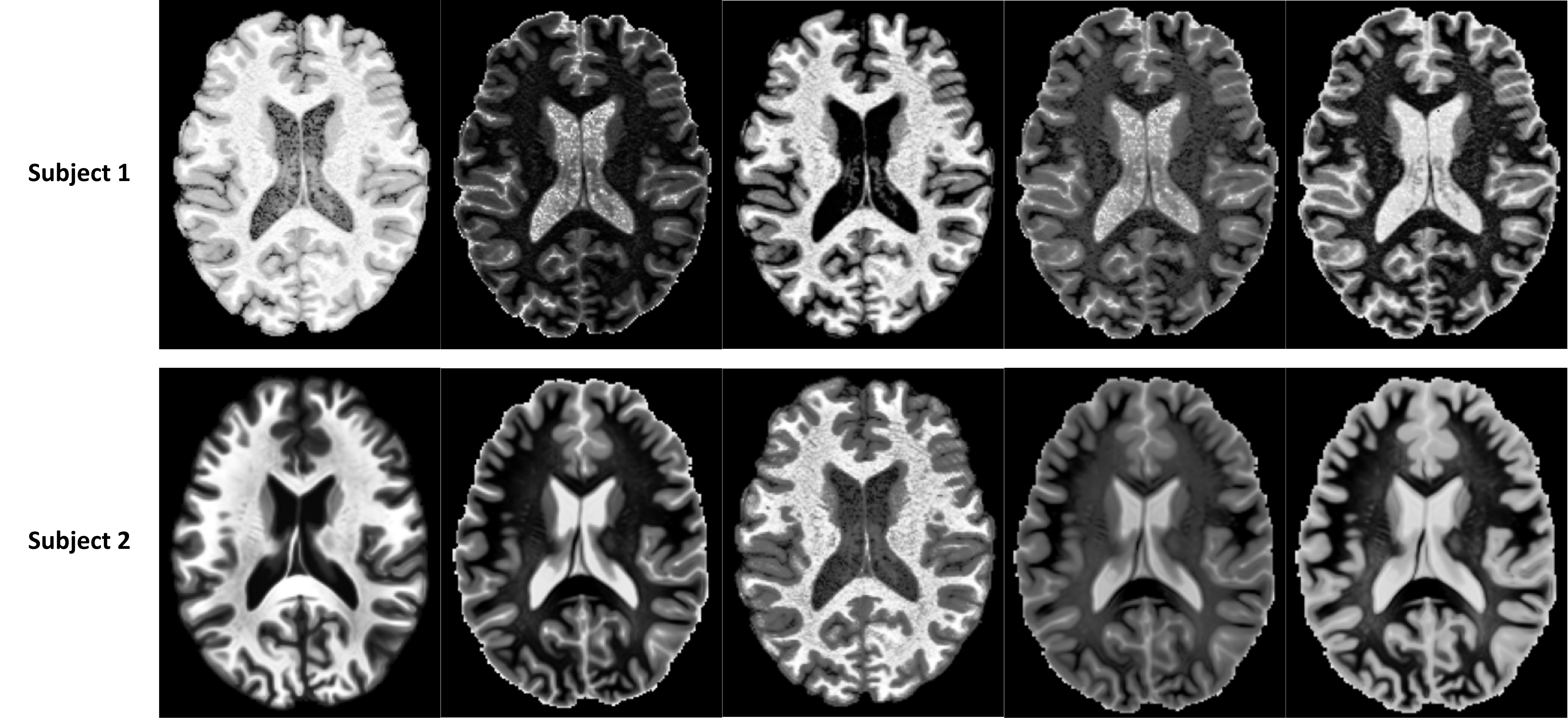}
\caption{Axial mid-slices of augmented samples generated using the RC-based contrast augmentation method with 1 $\times$ 1 $\times$ 1  convolution kernels.}
\label{RCKernel1}
\end{figure*}

\section{Random Convolution-Based Contrast Augmentation}

Figure \ref{RCModel} illustrates the model architecture used for random convolution (RC)-based contrast augmentation. The model consists of five non-linear blocks, each comprising an RC layer followed by a LeakyReLU activation. This cascaded design efficiently captures complex and non-linear intensity relationships across various MRI contrasts, generating diverse artificial contrast variations from a single scan. Additionally, Figure \ref{RCKernel1} presents axial mid-slices of augmented samples generated using the RC-based contrast augmentation scheme. These samples demonstrate the effectiveness of RC in simulating a wide range of artificial contrasts from a single input scan.

\begin{figure*}[tb]
\centering
\includegraphics[scale=0.39]{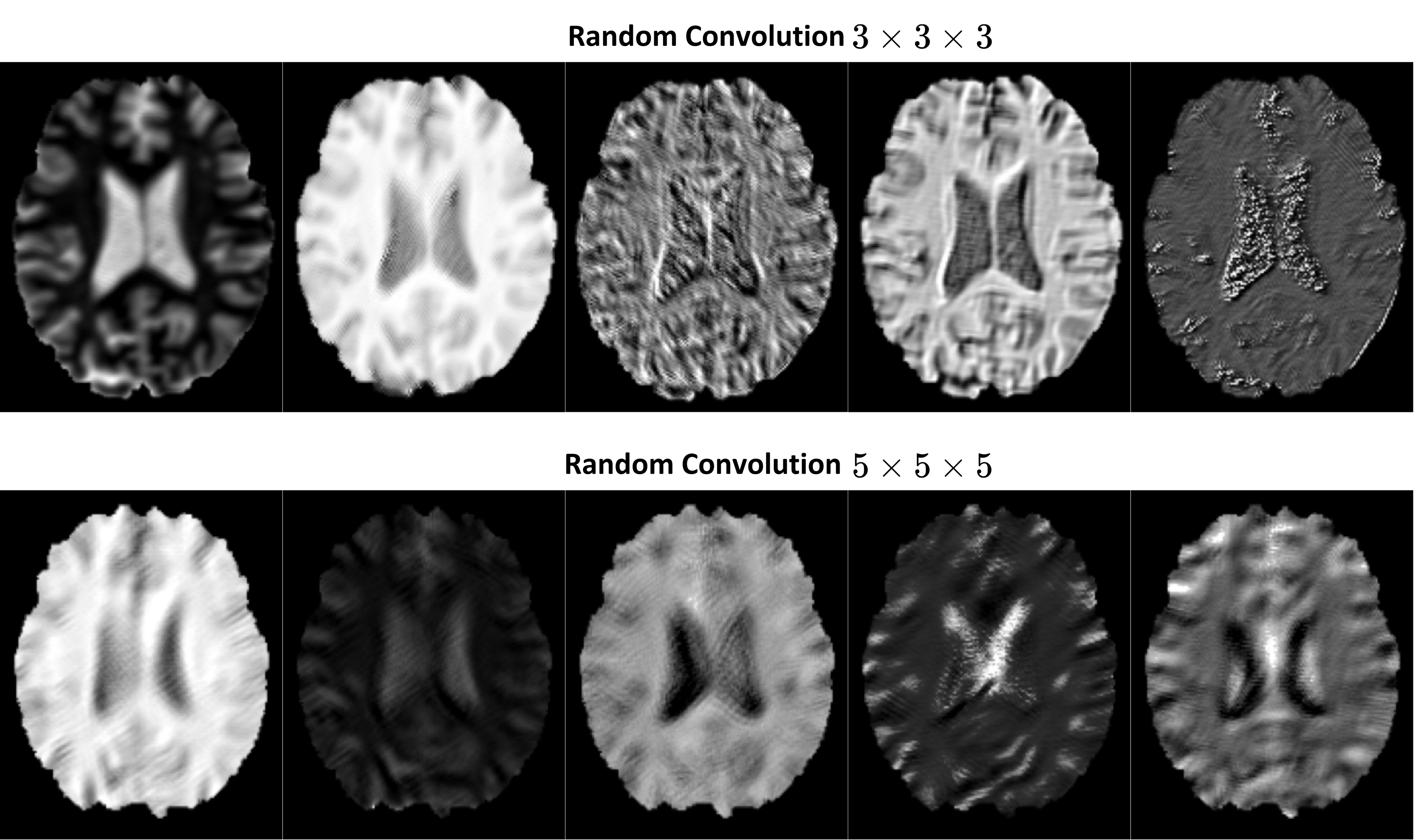}
\caption{Axial mid-slices of augmented samples generated using the RC-based contrast augmentation method with 3 $\times$ 3 $\times$ 3 and 5 $\times$ 5 $\times$ 5 convolution kernels, resulting in visible blurring effects.}
\label{RCKernel3_5}
\end{figure*}

To investigate the kernel size's impact on the RC output for contrast augmentation, we have implemented the kernel size of $3 \times 3 \times 3$ and $5 \times 5 \times 5$ in all non-linear blocks of our model. Figure \ref{RCKernel3_5} showcases axial mid-slices of augmented samples produced using the RC-based contrast augmentation method, incorporating $3 \times 3 \times 3$ and $5 \times 5 \times 5$ convolutions. Evidently, in these samples, the augmented outputs exhibit a noticeable blurring effect, which can adversely affect the performance of the DL models. In particular, this blurring compromises precise voxel-to-voxel correspondences for our task, thereby degrading the accuracy of anatomical landmark detection outcomes.

It is important to note that while the augmented scans with RC are inputs to the anatomical landmark detection model (\( \boldsymbol{f}(\cdot; \theta) \)), the calculated deformation field (Eq. \ref{DefFieldEq}) is applied to deform the scans before RC augmentation (Eq. \ref{SimLoss}) and the subsequent calculation of similarity and registration loss functions. This approach is based on the fact that RC does not alter the geometric properties of the scans but instead generates arbitrary contrast variations. This forces the model to predict landmarks independently of their contrasts. Consequently, this enables the use of a mono-modal loss function in Eq. \ref{SimLoss}, such as mean square error (MSE), while eliminating the need for computationally expensive metrics like mutual information (MI), normalized mutual information (NMI), or descriptors like modality independent neighborhood descriptor (MIND).

\section{Baselines}
It is important to note that we did not include the 3D U-Net as one of our baselines for direct landmark detection because it failed to converge and performed poorly on the publicly available test sets. This outcome was expected, as 3D U-Net typically has a much heavier parameter load compared to simpler architectures like the 3D supervised CNN we implemented. Given our limited labeled data (122 scans), the 3D U-Net struggled to converge effectively. Therefore, we opted to use a 3D CNN as the supervised learning baseline, which is more suited for scenarios with constrained datasets.

\section{Visual Comparison with ANTs and KeyMorph}
\begin{figure*}[tb] 
\centering
\includegraphics[scale=0.33]{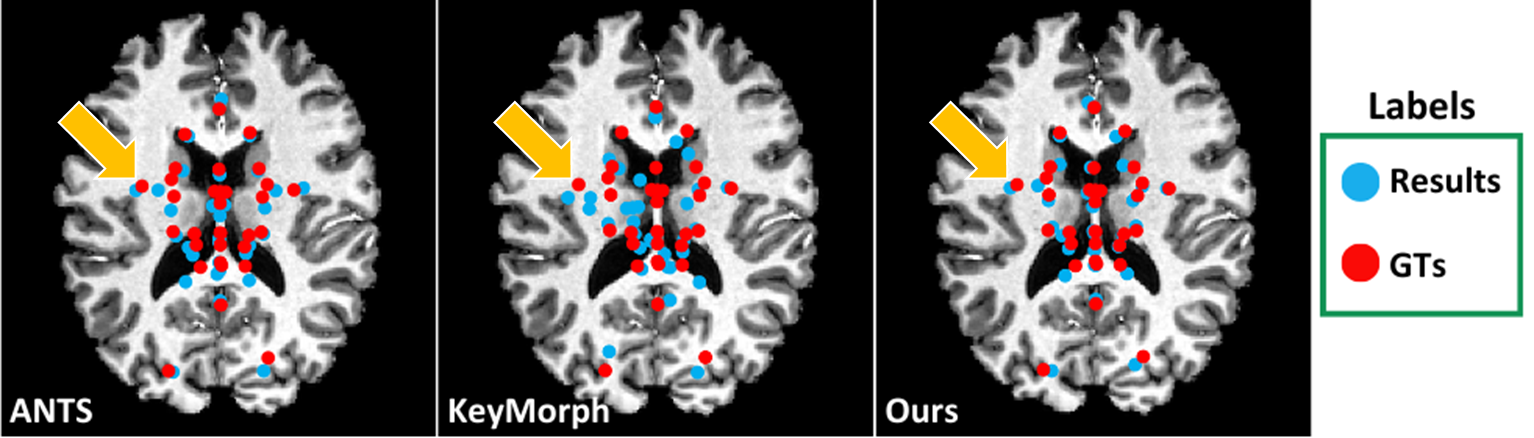}
\caption{Qualitative comparison of anatomical landmark detection results using our model, ANTs, and KeyMorph. Note that the landmarks shown are projections of 3D points in axial view for visual demonstration.}
\label{fig:LandmarkSam}
\end{figure*}

Samples of landmarks generated from our proposed model, ANTs, and KeyMorph for the same subject are shown in the axial view in Fig. \ref{fig:LandmarkSam} for comparison. Note that all landmarks are in 3D. For easy visualization, we show the 3D points projected in the 2D axial view while using a mid-axial MRI slice as a reference.

\section{Sensitivity to the Template Choice}
To assess the sensitivity of our method to the choice of template, we tested it using the widely adopted T1-weighted Colin27 atlas \cite{holmes1998enhancement} (a young, single-subject template) as an extreme alternative. The resulting MREs (in mm) were 5.87±4.02, 5.56±3.51, 4.92±3.12, and 5.38±3.32 for the SNSX, OASIS, HCP, and HCP-T2w datasets, respectively.
While the MREs are higher than using the ICBM152 (\textit{p}$<$0.05), they are on par with KeyMorph (512 KPs) with ICBM152 (\textit{p}$>$0.05). This is expected, as Fonov \textit{et al}. \cite{fonov2025understanding} shows that a stable group-average brain MRI template requires $\sim$160 subjects. We showed that ICBM152 didn't impact the results across age groups (\textit{p}$>$0.05). Following the suggestion of Fonov \textit{et al}. \cite{fonov2025understanding}, users have the flexibility to choose their own template and landmark protocol with our proposed method, as tagging a single template is not costly.

\section{Robustness to Pathological Brains}
While our current evaluation focuses on healthy subjects due to the availability of annotated data, assessing robustness for pathological brains is clinically important. As an indirect accuracy test, we evaluated CABLD for Parkinson's disease (PD) and Alzheimer's disease (AD) diagnosis (\textbf{Sec. 4.6}), confirming its sensitivity in detecting pathology-related landmark differences. For a direct test, we evaluated our method on 1.5T Gd-T1w MRIs of 36 PD patients from the London Health Sciences Center Parkinson’s disease \textbf{(LHSCPD)} dataset \cite{abbass2022application,taha2023magnetic}, featuring clinical scans with disease-related anatomical degeneration (e.g., atrophy) and additional domain shifts (unseen 1.5T field strength, low scan contrast/quality, and Gd MRI contrast agents). Our method achieved an \textbf{MRE=5.19±2.13mm}, not significantly different from the 40-90yo healthy group results (\textit{p}$>$0.05), with the same age range for PD patients. Finally, upon availability of AFIDs-compliant annotations, we will validate our method for other brain disorders.

\section{Computation Time}
Our method achieves an average inference time of 0.35±0.012s (GPU) and 6.42±0.20s (CPU), which is significantly faster than ANTs (MI: 428.22±3.14s, CC: 380.62±0.72s, CPU), and also faster than KeyMorph (10.12±0.22s CPU, 0.54±0.01s GPU) and BrainMorph (180.14±1.99s CPU, 1.24±0.30s GPU). These demonstrate the computational efficiency of our framework for large-scale and time-sensitive applications.

\end{document}